\documentclass[english]{article}
\usepackage[T1]{fontenc}
\usepackage[latin9]{inputenc}
\usepackage{geometry}
\usepackage{amsmath}
\usepackage{amssymb}
\usepackage{graphicx}
\usepackage{esint}

\makeatletter

\providecommand{\tabularnewline}{\\}

\usepackage{babel}

\usepackage{babel}
\IfFileExists{lmodern.sty}{\usepackage{lmodern}}{}

\usepackage{babel}

\usepackage{babel}

\usepackage{babel}

\@ifundefined{showcaptionsetup}{}{%
 \PassOptionsToPackage{caption=false}{subfig}}
\usepackage{subfig}
\makeatother

\usepackage{babel}
\begin{document}

\title{Diphoton decay for a 750 GeV scalar boson in a $SU(6)\otimes U(1)_{X}$
model}

\author{S. F. Mantilla$\thanks{e-mail:sfmantillas@unal.edu.co}$, R. Martinez$\thanks{e-mail:remartinezm@unal.edu.co}$,
F. Ochoa$\thanks{e-mail:faochoap@unal.edu.co}$, C. F. Sierra$\thanks{e-mail:cfsierraf@unal.edu.co}$.}

\date{\textit{Departamento de F\'{i}sica, Universidad Nacional de Colombia,
Ciudad Universitaria, K. 45 No. 26-85, Bogotá D.C., Colombia}}
\maketitle
\begin{abstract}
We propose a new $SU(6)\otimes U(1)_{X}$ GUT model free from anomalies,
with a 750 GeV scalar candidate which can decay into two photons,
compatible with the recent diphoton signal reported by ATLAS and CMS
collaborations. This model gives masses to all fermions and may explain
the 750GeV signal through one loop decays to $\gamma\gamma$ with
charged vector and charged Higgs bosons, as well as up- and electron-like
exotic particles that arise naturally from the condition of cancellation
of anomalies of the $SU(6)\otimes U(1)_{X}$ group. We obtain, for
different width approximations, allowed mass regions from 900 GeV
to 3 TeV for the exotic up-like quark, in agreement with ATLAS and
CMS collaborations data. 
\end{abstract}

\section{Introduction}

Recently the ATLAS and CMS collaborations reported a diphoton signal
excess with invariant mass of 750 GeV \cite{CMS750,ATLAS750} which
has been the subject of many interpretations in the literature using
different extensions of the standard model (SM) \cite{key-3,key-4,key-5,key-6,key-7,key-8,key-9,key-10,key-11}.
In this work, we consider the $SU(6)\otimes U(1)_{X}$ extension proposed
in \cite{Martinez SU6} in the framework of the flipped $SU(6)$ models
\cite{Cecotti} as a feasible model that may explain the diphoton
excess. These kind of flipped models have very interesting features.
First, by requiring a high breaking scale ($\sim10^{17}$ GeV) for
the flipped $SU(6)$ and its $SU(5)$ subgroup \cite{Panagiotakopoulos}
the proton decay problem can be avoid. Second, they are able to solve
the doublet-triplet splitting problem through the pseudo-Goldstone
mechanism as in $SU(6)$ \cite{Inoue,SU6 Barbieri} and $[SU(3)]^{3}$
\cite{Dvali}. Also, they provide unification of gauge couplings as
in the flipped $SU(5)$ model \cite{Barr,Antoniadis}. Finally, these
models may develop see-saw masses compatible with the phenomenological
active neutrinos \cite{Ellis flipped,Qaisar} if one singlet heavy
state is introduced.

The $SU(6)\otimes U(1)_{X}$ extension considered here contains the
$SU(3)_{C}\otimes SU(3)_{L}\otimes U(1)_{X}$ model (hereafter 331
model) \cite{Pleitez,Frampton,Long,Martinez 331} as a subgroup that
allow us address the observed diphoton excess through new exotic charged
Higgs bosons into the loop at the TeV scale. In the flipped model,
the $U(1)_{X}$ symmetry changes the exotic down type quark (charge
$-1/3$) by an up type quark (charge $2/3$) in the multiplets, which
increases the coupling with photons and gluons into the loop, resulting
in a significantly enhanced $pp\rightarrow\gamma\gamma$ cross section,
compatible with the reported data.

The 331 model can be embedded into the grand unified group $SU(6)\otimes U(1)_{X}$
with the following spontaneous symmetry breaking (SSB) chain: 
\begin{equation}
\begin{aligned}SU(6)\otimes U(1)_{X}\overset{{\mathbf{\Phi}}}{\xrightarrow{\hspace*{0.5cm}}}SU(3)_{C}\otimes SU(3)_{L}\otimes U(1)_{X}\overset{H_{1},H_{S}}{\xrightarrow{\hspace*{1cm}}}\\
SU(3)_{C}\otimes SU(2)_{L}\otimes U(1)_{Y}\overset{H_{2},H_{3},H_{S}}{\xrightarrow{\hspace*{1.5cm}}}SU(3)_{C}\otimes U(1)_{Q}
\end{aligned}
\label{eq:ssb chain}
\end{equation}
which are mediated by the five Higgs fields $\Phi$, $H_{1}$, $H_{2}$
, $H_{3}$ and $H_{S}$ in the $\mathbf{35}$, $\bar{\mathbf{6}}$,
$\bar{\mathbf{6}}$ ,$\bar{\mathbf{15}}$ and $\bar{\mathbf{15}}$
representations, respectively. From the mixing of the real components
of the fields $H_{1}$ and $H_{S}$ we will obtain two real scalar
fields, our candidate for the 750 GeV signal ($\xi$), and the other
at the TeV scale ($\xi'$).

This paper is organized as follows. In section 2, we show the particle
content of a $SU(6)\otimes U(1)_{X}$ model as an anomaly free theory
which contains the 331, 321 and 31 subgroups and their SSB scheme.
We describe the Yukawa Lagrangian showing that four Higgs fields are
sufficient to give masses to all fermions. We also show the most general
Higgs potential terms compatible with the symmetries and identify
the relevant quartic couplings that will induce the process $pp\rightarrow\xi\rightarrow\gamma\gamma$.
Section 3 is devoted to explore allowed regions consistent with the
reported cross section of the 750 GeV signal. Finally, in section
4, we summarize our conclusions.

\section{$SU(6)\otimes U(1)_{X}$ model}

$SU(6)\otimes U(1)_{X}$ strong-electroweak models provide us a new
framework which contains 331 and SM models for one family of fermions
as effective low energy field theories. In order to include the three
families we consider replicas of the first family as in the SM. Below,
we describe some remarkable properties of these models. 
\begin{itemize}
\item The cancellation of the $[SU(6)]^{3}$, $[SU(6)]^{2}U(1)_{X}$, $[\mathrm{Grav}]^{2}U(1)_{X}$
and $[U(1)_{X}]^{3}$ chiral anomaly equations, shown in reference
\cite{Martinez SU6}, provide us a set of multiplets with non-trivial
$U(1)_{X}$ charges which are family independent. We require two sextets
$\mathbf{\bar{6}}$, one antisymmetric $\mathbf{15}$ multiplet and
three singlets with charges $X_{\mathbf{\bar{6}}}=-2/3$, $X_{\mathbf{15}}=1/3$
and $X_{\mathbf{1}}=1$, respectively. 
\item The symmetry breaking $SU(6)\rightarrow SU(3)_{C}\otimes SU(3)_{L}\otimes U(1)_{Y'}$
gives us the following branching rules: 
\begin{align}
\bar{\mathbf{6}} & =(\bar{\mathbf{3}}_{C}\otimes\mathbf{1}_{L})_{1/3}\oplus(\mathbf{1}_{C}\otimes\mathbf{3}_{L})_{-1/3},\\
\mathbf{15} & =(\bar{\mathbf{3}}_{C}\otimes\mathbf{1}_{L})_{2/3}\oplus(\mathbf{1}_{L}\otimes\mathbf{3}_{L})_{-2/3}\oplus(\mathbf{3}_{C}\otimes\bar{\mathbf{3}}_{L})_{0},
\end{align}
where $(\mathbf{n}\otimes\mathbf{m})_{Y'}$ are tensorial products
of $\mathbf{n}$ $SU(3)_{C}$ multiplet with $\mathbf{m}$ $SU(3)_{L}$
multiplets and $Y'$ corresponds to the $U(1)_{Y'}$ quantum number
normalized as $2Y'/\sqrt{3}$, where: 
\begin{equation}
Y'=\frac{1}{2\sqrt{3}}\;\mathrm{diag}\left(\begin{array}{cccccc}
+1 & +1 & +1 & -1 & -1 & -1\end{array}\right).
\end{equation}
This gives us the following multiplets for the first family: 
\begin{equation}
\psi_{L}=\left(\begin{array}{c}
u_{}^{c}\\
u_{}^{c}\\
u_{}^{c}\\
\nu_{e}\\
e^{-}\\
E^{-}
\end{array}\right)_{L}\qquad\chi_{L}=\left(\begin{array}{c}
U_{}^{c}\\
U_{}^{c}\\
U_{}^{c}\\
N_{E}\\
E_{1}^{-}\\
E_{2}^{-}
\end{array}\right)_{L}\qquad\Psi_{L}^ {}=\left(\begin{array}{cccccc}
0 & d_{}^{c} & -d_{}^{c} & d_{} & u_{} & U_{}\\
-d_{}^{c} & 0 & d_{}^{c} & d_{} & u_{} & U_{}\\
d_{}^{c} & -d_{}^{c} & 0 & d_{} & u_{} & U_{}\\
-d_{} & -d_{} & -d_{} & 0 & \nu_{e1}^{c} & -N_{E1}^{c}\\
-u_{} & -u_{} & -u_{} & -\nu_{e}^{c} & 0 & E_{1}^{+}\\
-U_{} & -U_{} & -U_{} & N_{E}^{c} & -E_{1}^{+} & 0
\end{array}\right)_{L}
\end{equation}
\begin{equation}
\mathbf{1}:\qquad e_{L}^{+},\qquad E_{L}^{+},\qquad E_{2L}^{+},\qquad\nu_{SL}
\end{equation}
where $U$ is a new up-like quark, $E^{-}$, $E_{1}^{-}$ and $E_{2}^{-}$
are new exotic charged leptons and $\nu_{e1R}$, $N_{EL}$ and $N_{ER1}$
are new neutrinos. In order to obtain fermion mass hierarchies among
families, discrete symmetries can be introduced to obtain suitable
mass matrix ansatz. The additional sterile neutrino $\nu_{S}$ with
$X_{S}=0$ is necessary to produce see-saw mechanisms between neutrinos
\cite{Ma,Altarelli,carcamo}. 
\item The covariant derivatives for each type of multiplets are defined
as follows: 
\begin{eqnarray}
D_{\mu}\psi_{a} & = & \partial_{\mu}\psi_{a}+i\left(g_{6}A_{\mu}^{\alpha}(T_{\alpha})_{\;a}^{b}\right)\psi_{b}+ig_{X}\left(X_{\bar{\mathbf{6}}}\right)_{\;a}^{b}X'^{\mu}\psi_{b},\\
D_{\mu}\Psi^{ab} & = & \partial_{\mu}\Psi^{ab}-i\left(g_{6}A_{\mu}^{\alpha}(T_{\alpha})_{\;cd}^{ab}\right)\Psi^{cd}-ig_{X}\left(X_{\mathbf{15}}\right)_{\;cd}^{ab}X'^{\mu}\Psi^{cd},
\end{eqnarray}
where Latin indices run from 1 to 6, while Greek indices run from
1 to 35. The$\mathbf{15}$ generators are given by $(T_{\alpha})_{\;cd}^{ab}=(T_{\alpha})_{\;c}^{a}\delta_{\;d}^{b}+\delta_{\;c}^{a}(T_{\alpha})_{\;d}^{b}$. 
\item Gauge bosons are described by the $\mathbf{35}=\bar{\mathbf{35}}$
adjoint representation which obey the branching rule 
\begin{equation}
\begin{aligned}\mathbf{35}=(\mathbf{8}\otimes\mathbf{1})_{0}\oplus(\mathbf{1}\otimes\mathbf{8})_{0}\oplus(\mathbf{3}\otimes\bar{\mathbf{3}})_{2/3}\oplus(\bar{\mathbf{3}}\otimes\mathbf{3})_{-2/3}\oplus(\mathbf{1}\otimes\mathbf{1})_{0}\end{aligned}
,
\end{equation}
where $(\mathbf{8}\otimes\mathbf{1})_{0}$ are identified as QCD gluons;
$(\mathbf{1}\otimes\mathbf{8})_{0}$ are electro-weak gauge bosons
which contains $W^{\pm}_{\mu}$, $W^{\pm}_{3\mu}$, $W^{0}_{3\mu}$, $\overline{W}^{0}_{3\mu}$,
$A^{3}_{\mu}$ and $A^{8}_{\mu}$ bosons; $(\mathbf{1}\otimes\mathbf{1})_{0}$ is a
neutral boson $B_{Y'{\mu}}$ from the $U(1)_{Y'}$ symmetry, and $(\mathbf{3}\otimes\bar{\mathbf{3}})_{2/3}$
and $(\bar{\mathbf{3}}\otimes\mathbf{3})_{-2/3}$ are new leptoquark
bosons: $X_{\mu}$ with electric charge $2/3$, and $Y_{1{\mu}}$ and $Y_{{\mu}2}$
with electric charge $-1/3$, which induces quark-lepton interchange
processes. Their corresponding multiplet is: 
\begin{equation}
\mathbf{A}=\frac{1}{\sqrt{2}}\left(\begin{array}{cccccc}
G_{\;\;1}^{1} & G_{\;\;2}^{1} & G_{\;\;3}^{1} & X^{1c} & Y_{1}^{1c} & Y_{2}^{1c}\\
G_{\;\;1}^{2} & G_{\;\;2}^{2} & G_{\;\;3}^{2} & X^{2c} & Y_{1}^{2c} & Y_{2}^{2c}\\
G_{\;\;1}^{3} & G_{\;\;2}^{3} & G_{\;\;3}^{3} & X^{3c} & Y_{1}^{3c} & Y_{2}^{3c}\\
X^{1} & X^{2} & X^{3} & D_{1} & W^{+} & W_{3}^{+}\\
Y_{1}^{1} & Y_{1}^{2} & Y_{1}^{3} & W^{-} & D_{2} & W_{3}^{0}\\
Y_{2}^{1} & Y_{2}^{2} & Y_{2}^{3} & W_{3}^{-} & \overline{W}_{3}^{0} & D_{3}
\end{array}\right)
\end{equation}
where $G_{\;\;1}^{1}+G_{\;\;2}^{2}+G_{\;\;3}^{3}=0$. $D_{1}=A^{3}/\sqrt{2}+A^{8}/\sqrt{6}$,
$D_{2}=-A^{3}/\sqrt{2}+A^{8}/\sqrt{6}$ and $D_{3}=-\sqrt{2}A^{8}/\sqrt{3}$
are the diagonal $SU(3)_{L}$ gauge fields. In addition, there is
a new electrically neutral vector boson $X_{\mu}$ from $U(1)_{X}$ symmetry.
In total, the $SU(6)\otimes U(1)_{X}$ group has 36 gauge bosons:
eight gluons, eight electroweak bosons, eighteen leptoquark bosons
and two electrically neutral bosons. 
\item Electric charge are constructed using all diagonal generators of $SU(6)\otimes U(1)_{X}$:
\begin{equation}
{Q}=aT_{3}+\frac{2b}{\sqrt{3}}T_{8}+\frac{2c}{\sqrt{6}}T_{15}+\frac{2d}{\sqrt{10}}T_{24}+\frac{2e}{\sqrt{15}}T_{35}+XI_{6},
\end{equation}
where the $a,b,c,d,e$ and $f$ constants are fixed such that the
electric charge match with each charge from the multiplets. We find
\begin{equation}
{Q}=T_{3L}-\frac{1}{2\sqrt{6}}T_{15}-\frac{1}{2\sqrt{10}}T_{24}+\frac{2}{\sqrt{15}}T_{35}+XI_{6}=T_{3L}+\frac{Y}{2},
\end{equation}
where $Y$ is the usual hypercharge operator of the SM. 
\item The fermions contained in the model have the charges listed in Table
\ref{tab:Quantum-numbers}.

\begin{table}[h]
\centering %
\begin{tabular}{ccccc}
\hline 
\multicolumn{5}{c}{Left-handed}\tabularnewline
\hline 
 & $T_{3L}$  & $X$  & $Y$  & $Q$ \tabularnewline
\hline 
\hline 
$u$  & $+1/2$  & $+1/3$  & $+1/3$  & $+2/3$ \tabularnewline
\hline 
$d$  & $-1/2$  & $+1/3$  & $+1/3$  & $-1/3$ \tabularnewline
\hline 
$U$  & $0$  & $+1/3$  & $+4/3$  & $+2/3$ \tabularnewline
\hline 
\hline 
$\nu_{e}$  & $+1/2$  & $-2/3$  & $-1$  & $0$ \tabularnewline
\hline 
$e^{-}$  & $-1/2$  & $-2/3$  & $-1$  & $-1$ \tabularnewline
\hline 
\hline 
$N_{E}$  & $+1/2$  & $-2/3$  & $-1$  & $0$ \tabularnewline
\hline 
$E_{1}^{-}$  & $-1/2$  & $-2/3$  & $-1$  & $-1$ \tabularnewline
\hline 
\hline 
$E^{-}$  & $0$  & $-2/3$  & $-2$  & $-1$ \tabularnewline
\hline 
$E_{2}^{-}$  & $0$  & $-2/3$  & $-2$  & $-1$ \tabularnewline
\hline 
\end{tabular}$\quad$%
\begin{tabular}{ccccc}
\hline 
\multicolumn{5}{c}{Right-handed}\tabularnewline
\hline 
 & $T_{3L}$  & $X$  & $Y$  & $Q$ \tabularnewline
\hline 
\hline 
$u$  & $0$  & $+2/3$  & $+4/3$  & $+2/3$ \tabularnewline
\hline 
$d$  & $0$  & $+1/3$  & $-2/3$  & $-1/3$ \tabularnewline
\hline 
$U$  & $0$  & $+2/3$  & $+4/3$  & $+2/3$ \tabularnewline
\hline 
\hline 
$\nu_{e1}$  & $0$  & $-1/3$  & $0$  & $0$ \tabularnewline
\hline 
$e^{-}$  & $0$  & $-1$  & $-2$  & $-1$ \tabularnewline
\hline 
\hline 
$N_{E1}$  & $+1/2$  & $-1/3$  & $-1$  & $0$ \tabularnewline
\hline 
$E_{1}^{-}$  & $-1/2$  & $-1/3$  & $-1$  & $-1$ \tabularnewline
\hline 
\hline 
$E^{-}$  & $0$  & $-1$  & $-2$  & $-1$ \tabularnewline
\hline 
$E_{2}^{-}$  & $0$  & $-1$  & $-2$  & $-1$ \tabularnewline
\hline 
\end{tabular}\caption{Quantum numbers for the fermionic sector of the model. \label{tab:Quantum-numbers}}
\end{table}

\item The scalar sector is introduced to obtain the correct SSB chain. The
two last symmetry breakings are fulfilled using two Higgs fields represented
by sextets $\bar{\mathbf{6}}$ with $X_{H_{1},H_{2}}=1/3$. The directions
of their VEV, $V_{1}$ and $v_{2}$, are selected to obtain electrically
neutral vacua. In addition, $V_{1}$ is at the TeV scale while $v_{2}$
is at the electroweak scale. Two additional Higgs fields represented
by $\mathbf{\bar{15}}$ multiplets with $X_{H_{3}}=-2/3$ and $X_{H_{S}}=1/3$
are introduced to give masses to down quarks and neutrinos, respectively.

The first SSB needs a Higgs field from $\mathbf{35}$ adjoint representation
with the following VEV: 
\begin{equation}
\left\langle \Phi\right\rangle _{0}=\mathcal{V}_{\mathrm{GUT}}\;\mathrm{diag}\left(\begin{array}{cccccc}
+1 & +1 & +1 & -1 & -1 & -1\end{array}\right),
\end{equation}
where $\mathcal{V}_{\mathrm{GUT}}\sim10^{17}$ GeV breaks the gauge
symmetry to $SU(3)_{C}\otimes SU(3)_{L}\otimes U(1)_{Y'}$ providing
masses to the leptoquark bosons. For the second and third SSBs, we
define the following Higgs scalar multiplets: 
\begin{equation}
H_{1}=\left(\begin{array}{c}
\phi_{1}^{1/3}\\
\phi_{1}^{1/3}\\
\phi_{1}^{1/3}\\
\varphi_{1}^{+}\\
\varphi_{1}^{0}\\
\frac{\xi_{1}+V_{1}+i\zeta_{1}}{\sqrt{2}}
\end{array}\right),\qquad H_{2}=\left(\begin{array}{c}
\phi_{2}^{1/3}\\
\phi_{2}^{1/3}\\
\phi_{2}^{1/3}\\
\varphi_{2}^{+}\\
\frac{h_{2}+v_{2}+i\eta_{2}}{\sqrt{2}}\\
\varphi_{2}^{0}
\end{array}\right)\label{eq:H1.H2}
\end{equation}
\begin{equation}
H_{3}=\left(\begin{array}{cccccc}
0 & \phi_{3}^{2/3} & -\phi_{3}^{2/3} & \phi_{3}^{4/3} & \phi_{31}^{1/3} & \phi_{32}^{1/3}\\
-\phi_{3}^{2/3} & 0 & \phi_{3}^{2/3} & \phi_{3}^{4/3} & \phi_{31}^{1/3} & \phi_{32}^{1/3}\\
\phi_{3}^{2/3} & -\phi_{3}^{2/3} & 0 & \phi_{3}^{4/3} & \phi_{31}^{1/3} & \phi_{32}^{1/3}\\
-\phi_{3}^{4/3} & -\phi_{3}^{4/3} & -\phi_{3}^{4/3} & 0 & \phi_{3}^{+} & -\varphi_{3}^{+}\\
-\phi_{31}^{1/3} & -\phi_{31}^{1/3} & -\phi_{31}^{1/3} & -\phi_{3}^{+} & 0 & \frac{h_{3}+v_{3}+i\eta_{3}}{\sqrt{2}}\\
-\phi_{32}^{1/3} & -\phi_{32}^{1/3} & -\phi_{32}^{1/3} & \varphi_{3}^{+} & -\frac{h_{3}+v_{3}+i\eta_{3}}{\sqrt{2}} & 0
\end{array}\right),\label{eq:H3}
\end{equation}
\begin{equation}
H_{S}=\left(\begin{array}{cccccc}
0 & \phi_{S}^{-1/3} & -\phi_{S}^{-1/3} & \phi_{S}^{1/3} & \phi_{S1}^{-2/3} & \phi_{S2}^{-2/3}\\
-\phi_{S}^{-1/3} & 0 & \phi_{S}^{-1/3} & \phi_{S}^{1/3} & \phi_{S1}^{-2/3} & \phi_{S2}^{-2/3}\\
\phi_{S}^{-1/3} & -\phi_{S}^{-1/3} & 0 & \phi_{S}^{1/3} & \phi_{S1}^{-2/3} & \phi_{S2}^{-2/3}\\
-\phi_{S}^{1/3} & -\phi_{3}^{4/3} & -\phi_{3}^{4/3} & 0 & \frac{\xi_{S}+V_{S}-i\zeta_{S}}{\sqrt{2}} & -\frac{h_{S}+v_{S}-i\eta_{S}}{\sqrt{2}}\\
-\phi_{S1}^{-2/3} & -\phi_{S1}^{-2/3} & -\phi_{S1}^{-2/3} & -\frac{\xi_{S}+V_{S}-i\zeta_{S}}{\sqrt{2}} & 0 & \varphi_{S}^{+}\\
-\phi_{S2}^{-2/3} & -\phi_{S2}^{-2/3} & -\phi_{S2}^{-2/3} & \frac{h_{S}+v_{S}-i\eta_{S}}{\sqrt{2}} & -\varphi_{S}^{+} & 0
\end{array}\right),\label{eq:HS}
\end{equation}
where $V_{1},V_{S}\gg v_{2},v_{3},v_{S}\sim246\mathrm{GeV}$. In this
way, the SSB chain is given by Eq.(\ref{eq:ssb chain}).

\item Vector boson masses: there are two electroweak SSBs in the low-energy
$SU(3)_{L}\otimes U(1)_{X}$ model, the first $V$ at TeV and the
second at $v$ GeV scale. After the TeV SSB $SU(3)_{L}\otimes U(1)_{X}\rightarrow SU(2)_{L}\otimes U(1)_{Y}$
the gauge bosons $A_{\mu}^{8}$ and $X_{\mu}$ mix them together into
the weak hypercharge boson $B_{\mu}$ and a new massive electrically
neutral gauge boson $Z_{X\mu}$ trough the following mixing matrix
with the mixing angle $\tan\theta_{X}=-\sqrt{3}{g}/{g_{X}}$, 
\begin{equation}
\left(\begin{array}{c}
B\\
Z_{X}
\end{array}\right)=\left(\begin{array}{cc}
\cos\theta_{X} & -\sin\theta_{X}\\
\sin\theta_{X} & \cos\theta_{X}
\end{array}\right)\left(\begin{array}{c}
A^{8}\\
-X
\end{array}\right).
\end{equation}
The new gauge coupling constant is the electroweak hypercharge $g'=g_{X}\sin\theta_{X}=-\sqrt{3}g\cos\theta_{X}$.
The gauge bosons $W_{3\mu}^{\pm}$, $W_{3\mu}^{0}$ and $\overline{W}_{3\mu}^{0}$
acquire the same mass $M_{W_{3}}$ which is related to $M_{Z_{X}}$
by $\sin\theta_{W}$ in the following way 
\begin{equation}
M_{W_{3}}=\frac{gV}{2}\qquad,\qquad M_{Z_{X}}=\frac{gV}{\sqrt{3}\sin\theta_{X}}=\frac{2}{\sqrt{3}}\frac{M_{W_{3}}}{\sin\theta_{X}}
\end{equation}
where $V^{2}=V_{1}^{2}+2V_{S}^{2}$. In addition, the $\xi$ gauge
couplings are given by 
\begin{equation}
\begin{split}\mathcal{L_{\xi VV}} & =\frac{\mathit{V}g^{2}}{3\sin^{2}\theta_{X}}\xi{Z_{X\mu}Z_{X}^{\mu}}
 + \frac{\mathit{V}{{\mathit{g}}^{2}}}{2}\xi\mathit{W_{3\mu}^{-}}\mathit{W_{3}^{+\mu}}
 + \frac{\mathit{V}{{\mathit{g}}^{2}}}{2}\xi\mathit{W_{3\mu}^{0}}\mathit{\overline{W}_{3}^{0\mu}}\\
 & =\frac{M_{Z_{1}}^{2}}{V}\xi{Z_{X\mu}Z_{X}^{\mu}} 
 + 2\frac{M_{W_{3}}^{2}}{V}\xi\mathit{W_{3\mu}^{-}}\mathit{W_{3}^{+\mu}}
 + 2\frac{M_{W_{3}}^{2}}{V}\xi\mathit{W_{3\mu}^{0}}\mathit{\overline{W}_{3}^{0\mu}}.
\end{split}
\end{equation}
Here $\xi$ does not couple to $B_{\mu}$ because $M_{B}=0$. Secondly,
for the GeV SSB $SU(2)_{L}\otimes U(1)_{Y}\rightarrow U(1)_{Q}$ will
bring the well-known gauge boson mixing through the Weinberg angle
$\tan\theta_{W}={g'}/{g}$ 
\begin{equation}
\left(\begin{array}{c}
Z_{W}\\
A
\end{array}\right)=\left(\begin{array}{cc}
\cos\theta_{W} & -\sin\theta_{W}\\
\sin\theta_{W} & \cos\theta_{W}
\end{array}\right)\left(\begin{array}{c}
A^{3}\\
B
\end{array}\right).
\end{equation}
The new gauge coupling constant is the electromagnetic charge $e=g_{2}\sin\theta_{W}=g_{Y}\cos\theta_{W}$
and the new gauge boson masses are 
\begin{equation}
M_{W_{3}^{+}}=\frac{g}{2}\sqrt{V^{2}+v^{2}},\qquad M_{W_{3}^{0}}=\frac{gV}{2},\qquad M_{W^{+}}=\frac{gv}{2}
\end{equation}
where $v^{2}=v_{2}^{2}+2v_{3}^{2}+2v_{S}^{2}=(246\mathrm{\,GeV})^{2}$.
In addition, $Z_{W\mu}$ and $Z_{X\mu}$ acquire the following masses
\begin{equation}
M_{Z_{W}}=\frac{gv}{2\cos\theta_{W}}=\frac{M_{W}}{\cos\theta_{W}},\qquad M_{Z_{X}}=\frac{2}{\sqrt{3}}\frac{M_{W_{3}}}{\sin\theta_{X}}.
\end{equation}

There is an additional gauge boson mixing between the two neutral
$Z_{W}$ and $Z_{X}$ through the mixing angle $\tan\theta_{Z}\propto{v^{2}}/{V^{2}}$,
\begin{equation}
\left(\begin{array}{c}
Z\\
Z'
\end{array}\right)=\left(\begin{array}{cc}
\cos\theta_{Z} & \sin\theta_{Z}\\
-\sin\theta_{Z} & \cos\theta_{Z}
\end{array}\right)\left(\begin{array}{c}
Z_{W}\\
Z_{X}
\end{array}\right)
\end{equation}
obtaining the physical gauge boson masses $Z$ and $Z'$, 
\begin{equation}
M_{Z}\approx M_{Z_{W}}\sqrt{1+\mathcal{O}\left(\frac{v^{2}}{V^{2}}\right)},\qquad M_{Z'}\approx M_{Z_{X}}\sqrt{1-\mathcal{O}\left(\frac{v^{2}}{V^{2}}\right)}
\end{equation}

\end{itemize}

\subsection{Yukawa Lagrangian}

The Yukawa Lagrangian that describes interactions between the Higgs
and the fermion sector is the following 
\begin{alignat}{1}
-\mathcal{L}_{\mathrm{Yukawa}}= & \sum_{i=1,2}(\psi_{La}^{\mathrm{T}}\hat{C}h_{\psi i}\Psi_{L}^{ab}H_{ib}+\psi_{La}^{\mathrm{T}}\hat{C}h_{\psi ei}H_{i}^{a}e_{L}^{+}+\psi_{La}^{\mathrm{T}}\hat{C}h_{\psi Ei}H_{i}^{a}E_{L}^{+}+\psi_{La}^{\mathrm{T}}\hat{C}h_{\psi E_{2}i}H_{i}^{a}E_{2L}^{+}\nonumber \\
 & +\;\chi_{La}^{\mathrm{T}}\hat{C}h_{\chi i}\Psi_{L}^{ab}H_{ib}+\chi_{La}^{\mathrm{T}}\hat{C}h_{\chi ei}H_{i}^{a}e_{L}^{+}+\chi_{La}^{\mathrm{T}}\hat{C}h_{\chi Ei}H_{i}^{a}E_{L}^{+}+\chi_{La}^{\mathrm{T}}\hat{C}h_{\chi E_{2}i}H_{i}^{a}E_{2L}^{+})+\nonumber \\
 & +\;h_{3}\epsilon_{abcdef}\Psi_{L}^{\mathrm{T}ab}\hat{C}\Psi_{L}^{dc}H_{3}^{ef}+{\Psi_{L}^{\mathrm{T}}}^{ab}\hat{C}h_{S}{H_{S}}_{ab}\nu_{SL}+\nu_{SL}^{T}\hat{C}M_{S}\nu_{SL}+h.c,\label{eq:Yukawa}
\end{alignat}
where $a,\cdots,f=1,\cdots,6$. The terms $\psi_{La}^{\mathrm{T}}\hat{C}h_{\psi i}\Psi_{L}^{ab}H_{ib}$
and $\chi_{La}^{\mathrm{T}}\hat{C}h_{\chi i}\Psi_{L}^{ab}H_{ib}$
give masses to up quarks and leptons at the order of $\left\langle H_{i}\right\rangle $
and with Yukawa coupling constants $h_{\psi i}$ and $h_{\chi i}$.
Terms containing couplings with singlet leptons, as for example $\psi_{La}^{\mathrm{T}}\hat{C}h_{\psi ei}H_{i}^{a}e_{L}^{+}$,
give masses to $e_{L}^{+}$, $E_{L}^{+}$ and $E_{2L}^{+}$ with coupling
constants $h_{\psi ei}$, $h_{\psi Ei}$ and $h_{\psi E_{2}i}$, respectively.
The term that contains couplings with the scalar fields $H_{3}$,
gives masses to down quarks with coupling constant $h_{3}$. The last
two terms in Eq. (\ref{eq:Yukawa}) induce see-saw neutrino masses
\cite{Wyler,Mohapatra,Hooft,Erika}, where $\nu_{SL}$ is a Majorana
neutrino of mass $M_{S}$ which is fixed to give a light neutrino
$\nu_{e}$, with mass at the order of eV.

First, for up quarks, we obtain the following non-diagonal mass matrix,
\begin{equation}
M_{uU}^{0}=\left(\begin{array}{cc}
h_{\psi2}v_{2} & h_{\psi1}V_{1}\\
h_{\chi2}v_{2} & h_{\chi1}V_{1}
\end{array}\right)
\end{equation}
By diagonalizing the symmetric matrix $\left(M_{uU}^{0}\right)^{\mathrm{T}}M_{uU}^{0}$,
we obtain the following masses 
\begin{eqnarray}
m_{U} & \approx & \sqrt{\frac{h_{\chi1}^{2}+h_{\psi1}^{2}}{2}}V_{1},\nonumber \\
m_{u} & \approx & \frac{h_{\chi1}h_{\psi2}-h_{\psi1}h_{\chi2}}{\sqrt{h_{\chi1}^{2}+h_{\psi1}^{2}}}v_{2}.
\end{eqnarray}
Charged leptons acquire masses through the following mass matrix in
the $(e^{-},E^{-},E_{1}^{-},E_{2}^{-})$ basis, 
\begin{equation}
M_{E}=\left(\begin{array}{cccc}
h_{\psi e2}v_{2} & h_{\psi E2}v_{2} & h_{\psi1}V_{1} & h_{\psi E_{2}2}v_{2}\\
h_{\psi e1}V_{1} & h_{\psi E1}V_{1} & -h_{\psi2}v_{2} & h_{\psi E_{2}1}V_{1}\\
h_{\chi e2}v_{2} & h_{\chi E2}v_{2} & h_{\chi1}V_{1} & h_{\chi E_{2}2}v_{2}\\
h_{\chi e1}V_{1} & h_{\chi E1}V_{1} & -h_{\chi2}v_{2} & h_{\chi E_{2}1}V_{1}
\end{array}\right),
\end{equation}
while neutrinos acquire masses through the following mass matrix in
the $(\nu_{e},\nu_{e1}^{c},N_{E},N_{E1}^{c},\nu_{S})$ basis, 
\begin{equation}
M_{N}=\left(\begin{array}{ccccc}
0 & h_{\psi2}v_{2} & 0 & -h_{\psi1}V_{1} & 0\\
h_{\psi2}v_{2} & 0 & h_{\chi2}v_{2} & 0 & h_{S}V_{S}\\
0 & h_{\chi2}v_{2} & 0 & -h_{\chi1}V_{1} & 0\\
-h_{\psi1}V_{1} & 0 & -h_{\chi1}V_{1} & 0 & h_{S}v_{S}\\
0 & h_{S}V_{S} & 0 & h_{S}v_{S} & M_{S}
\end{array}\right).
\end{equation}
Finally, the down quark acquire mass proportional to $\langle H_{3}\rangle=v_{3}$,
\begin{equation}
m_{d}=\sqrt{2}h_{3}v_{3}.
\end{equation}

\subsection{Higgs potential}

The interactions between the four Higgs bosons are described by the
following scalar potential, 
\begin{alignat}{1}
V(H_{1},H_{2},H_{3},H_{S},\Phi)= & -\mu_{1}^{2}H_{1}^{\dagger}H_{1}-\mu_{2}^{2}H_{2}^{\dagger}H_{2}-\mu_{3}^{2}\mathrm{Tr}(H_{3}^{\dagger}H_{3})-\mu_{S}^{2}\mathrm{Tr}(H_{S}^{\dagger}H_{S})^{2}-\mu_{\Phi}^{2}\mathrm{Tr}(\Phi)^{2}+\nonumber \\
 & \lambda_{1}(H_{1}^{\dagger}H_{1})^{2}+\lambda_{2}(H_{2}^{\dagger}H_{2})^{2}+\lambda{}_{3}\mathrm{Tr}(H_{3})^{4}+\lambda'_{3}(\mathrm{Tr}(H_{3})^{2})^{2}+\nonumber \\
 & \lambda{}_{S}\mathrm{Tr}(H_{S})^{4}+\lambda'_{S}(\mathrm{Tr}(H_{S})^{2})^{2}+\lambda{}_{\Phi}\mathrm{Tr}(\Phi)^{4}+\lambda'_{\Phi}(\mathrm{Tr}(\Phi)^{2})^{2}+\nonumber \\
 & \lambda_{12}(H_{1}^{\dagger}H_{1})(H_{2}^{\dagger}H_{2})+\lambda_{1S}(H_{1}^{\dagger}H_{1})\mathrm{Tr}(H_{S}^{\dagger}H_{S})^{2}+\lambda'_{1S}(H_{1}^{\dagger}(H_{S})^{2}H_{1})+\nonumber \\
 & \lambda_{13}(H_{1}^{\dagger}H_{1})\mathrm{Tr}(H_{3})^{2}+\lambda'_{13}(H_{1}^{\dagger}(H_{3})^{2}H_{1})+\lambda_{2S}(H_{2}^{\dagger}H_{2})\mathrm{Tr}(H_{S}^{\dagger}H_{S})^{2}+\nonumber \\
 & \lambda'_{2S}(H_{2}^{\dagger}(H_{S})^{2}H_{2})+\lambda_{23}(H_{2}^{\dagger}H_{2})\mathrm{Tr}(H_{3})^{2}+\lambda'_{23}(H_{2}^{\dagger}(H_{3})^{2}H_{2})+\label{eq:scalar-potential}\\
 & \lambda_{3S}(\mathrm{Tr}H_{3}^{2})(\mathrm{Tr}H_{S}^{2})+\lambda'_{3S}\mathrm{Tr}(H_{3}^{\dagger}H_{3}H_{S}^{\dagger}H_{S})+\lambda''_{3S}\mathrm{Tr}(H_{3}^{\dagger}H_{S}^{\dagger}H_{S}H_{3})+\nonumber \\
 & \lambda_{1\Phi}(H_{1}^{\dagger}H_{1})\mathrm{Tr}(H_{\Phi})^{2}+\lambda'_{1\Phi}(H_{1}^{\dagger}(H_{\Phi})^{2}H_{1})+\lambda_{2\Phi}(H_{2}^{\dagger}H_{2})\mathrm{Tr}(H_{\Phi})^{2}+\nonumber \\
 & \lambda'_{2\Phi}(H_{2}^{\dagger}(H_{\Phi})^{2}H_{2})+\lambda_{S\Phi}(\mathrm{Tr}H_{S}^{2})(\mathrm{Tr}\Phi^{2})+\lambda'_{S\Phi}\mathrm{Tr}(H_{S}\Phi H_{S}\Phi)+\nonumber \\
 & \lambda''_{S\Phi}\mathrm{Tr}(\Phi H_{S}H_{S}\Phi)+\lambda_{3\Phi}(\mathrm{Tr}H_{3}^{2})(\mathrm{Tr}\Phi^{2})+\lambda'_{3\Phi}\mathrm{Tr}(H_{3}\Phi H_{3}\Phi)+\nonumber \\
 & \lambda''_{3\Phi}\mathrm{Tr}(\Phi H_{3}H_{3}\Phi).\nonumber 
\end{alignat}

From $H_{2}$, $H_{3}$ and $H_{S}$ (equations \eqref{eq:H1.H2},
\eqref{eq:H3} and \eqref{eq:HS}) we obtain three electroweak Higgs
doublets which breaks $SU(2)_{L}\times U(1)_{Y}$ into $U(1)_{Q}$.
They are expressed as follow, 
\begin{equation}
H_{2_{EW}}=\left(\begin{array}{c}
\varphi_{2}^{+}\\
\frac{h_{2}+v_{2}+i\eta_{2}}{\sqrt{2}}
\end{array}\right),\qquad H_{3_{EW}}=\left(\begin{array}{c}
\varphi_{3}^{+}\\
\frac{h_{3}+v_{3}+i\eta_{3}}{\sqrt{2}}
\end{array}\right),\qquad H_{S_{EW}}=\left(\begin{array}{c}
\varphi_{S}^{+}\\
\frac{h_{S}+v_{S}+i\eta_{S}}{\sqrt{2}}
\end{array}\right)
\end{equation}
Thus, the model contains an effective three Higgs doublet model. For
the charged sector, $\varphi_{2}^{\pm}$, $\varphi_{3}^{\pm}$ and
$\varphi_{S}^{\pm}$ rotate into the Goldstone bosons $G^{\pm}$ associated
to $W^{\pm}$, and the physical charged Higgs bosons $H_{1}^{\pm}$
and $H_{2}^{\pm}$.

\subsection{Couplings with $\boldsymbol{\xi}$}

As we mentioned before, from the mixing terms between the real fields
$\xi_{1}$ in $H_{1}$ and $\xi_{S}$ in $H_{S}$, we obtain two real
scalar fields: our candidate for the 750 GeV signal $\xi$, and one
$\xi'$ at the TeV scale. In particular, we are interested in the
following trilinear terms coming from the quartic couplings between
the weak multiplets of the Higgs potential Eq.(\eqref{eq:scalar-potential}),

\begin{equation}
\begin{aligned}V_{\mathrm{trilinear}}=\lambda_{H_{1}}V\xi H_{1}^{+}H_{1}^{-}+\lambda_{H_{2}}V\xi H_{2}^{+}H_{2}^{-}+\lambda_{13}V\xi\phi_{3}^{+}\phi_{3}^{-}\end{aligned}
\label{trilinear-higgs}
\end{equation}
where $\lambda_{H_{1}}$ and $\lambda_{H_{2}}$ are linear combinations
of $\lambda_{12}$, $\lambda_{13}$ and $\lambda_{1S}$. The field
$\phi_{3}^{\pm}$ is a charged singlet Higgs-like boson coming from
the 15-dimensional representation $H_{3}$ in Eq.(\ref{eq:H3}). For
simplicity, we choose $\lambda_{H}=\lambda_{H_{1}}\sim\lambda_{H_{2}}\sim\lambda_{13}$.
Thus, the equation (\ref{trilinear-higgs}) becomes 
\begin{equation}
V_{\mathrm{trilinear}}=\lambda_{H}V\xi\left(H_{1}^{+}H_{1}^{-}+H_{2}^{+}H_{2}^{-}+\phi_{3}^{+}\phi_{3}^{-}\right).\label{trilinear}
\end{equation}

For the vector boson sector, for simplicity and following \cite{Cao}
we consider general interactions of the form

\begin{align}
g_{\gamma W_{3}W_{3}}=\kappa g_{ZWW}^{{\rm SM}},\quad & g_{\gamma H^{+}H^{-}}=\lambda\left(p_{1}-p_{2}\right)^{\mu}\label{eq:gammacouplings}
\end{align}

\begin{align}
g_{ZW_{3}W_{3}}=\kappa^{\prime}g_{ZWW}^{{\rm SM}}, & \quad g_{ZH^{+}H^{-}}=\lambda^{\prime}\left(p_{1}-p_{2}\right)^{\mu},\quad g_{ZH^{\text{+}}W_{3}^{-}}=\eta m_{W_{3}}g^{\mu\nu}.\label{eq:zcouplings}
\end{align}

In addition, from the fact that the scalar boson $\xi$ has not electroweak
isospin and hypercharge, its couplings with the $A_{\mu}^{3}$ and
$B_{\mu}$ gauge bosons are completely null, hence after the electroweak
SSB at the GeV scale the $\xi$ boson remains without interaction
with $A_{\mu}$ and $Z_{W\mu}$. Moreover, as the gauge bosons $Z_{W\mu}$
and $Z_{X\mu}$ mix them together into the physical gauge bosons $Z_{\mu}$
and $Z'_{\mu}$, there could be some interaction between $\xi$ and
$Z_{\mu}$. However, it is strongly suppressed by the $Z-$mixing
angle $\tan\theta_{Z}\propto{v^{2}}/{V^{2}}$.

Finally, for the fermionic sector, the flavor eigenstates of quarks
are related to their mass eigenstates trough the following mixing
matrix, 
\begin{equation}
\left(\begin{array}{c}
u'\\
c'\\
t'\\
U'\\
C'\\
T'
\end{array}\right)=\left(\begin{array}{ccc|ccc}
R_{uu} & R_{uc} & R_{ut} & R_{uU} & R_{uC} & R_{uT}\\
R_{cu} & R_{cc} & R_{ct} & R_{cU} & R_{cC} & R_{cT}\\
R_{tu} & R_{tc} & R_{tt} & R_{tU} & R_{tC} & R_{tT}\\
\hline R_{Uu} & R_{Uc} & R_{Ut} & R_{UU} & R_{UC} & R_{UT}\\
R_{Cu} & R_{Cc} & R_{Ct} & R_{CU} & R_{CC} & R_{CT}\\
R_{Tu} & R_{Tc} & R_{Tt} & R_{TU} & R_{TC} & R_{TT}
\end{array}\right)\left(\begin{array}{c}
u\\
c\\
t\\
U\\
C\\
T
\end{array}\right).
\end{equation}
The off-diagonal blocks mix SM and non-SM quarks. In particular, the
$t$ and $U$ quarks are related by 
\begin{equation}
\left(\begin{array}{c}
t'\\
U'
\end{array}\right)=\left(\begin{array}{c|c}
R_{tt} & R_{tU}\\
\hline R_{Ut} & R_{UU}
\end{array}\right)\left(\begin{array}{c}
t\\
U
\end{array}\right)
\end{equation}
where $R_{tt}=R_{UU}=\cos\theta_{tU}$ and $R_{Ut}=-R_{tU}=\sin\theta_{tU}\propto v_{2}/V_{1}$.
Since $v_{2}/V_{1}\ll1$ the mixing between $t$ and $U$ is small
resulting in the suppression of the coupling between $\xi$ and $t$.
In the same way the off-diagonal components of the mixing matrix in
Eq. (37) are proportional to $v_{2}/V_{1}$ resulting in the suppression
of these mixing terms splitting the up-quark sector in SM and non-SM
up-quarks.

\noindent \begin{center}
\begin{figure}[t]
\begin{centering}
\includegraphics[scale=0.35]{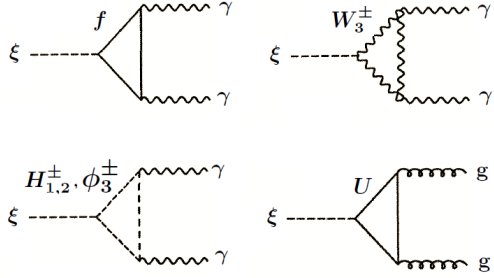} 
\par\end{centering}

\caption{One loop processes involved in the decay widths $\Gamma_{gg}$ and
$\Gamma_{\gamma\gamma}$. In this figure, $f=\{U,\:E^{-},\:E_{1}^{-},\:E_{2}^{-}\}$.
\label{fig:Loop-processes-involved}}
\end{figure}

\par\end{center}

\begin{center}
\begin{figure}[t]
\begin{centering}
\subfloat[]{\includegraphics[scale=0.8]{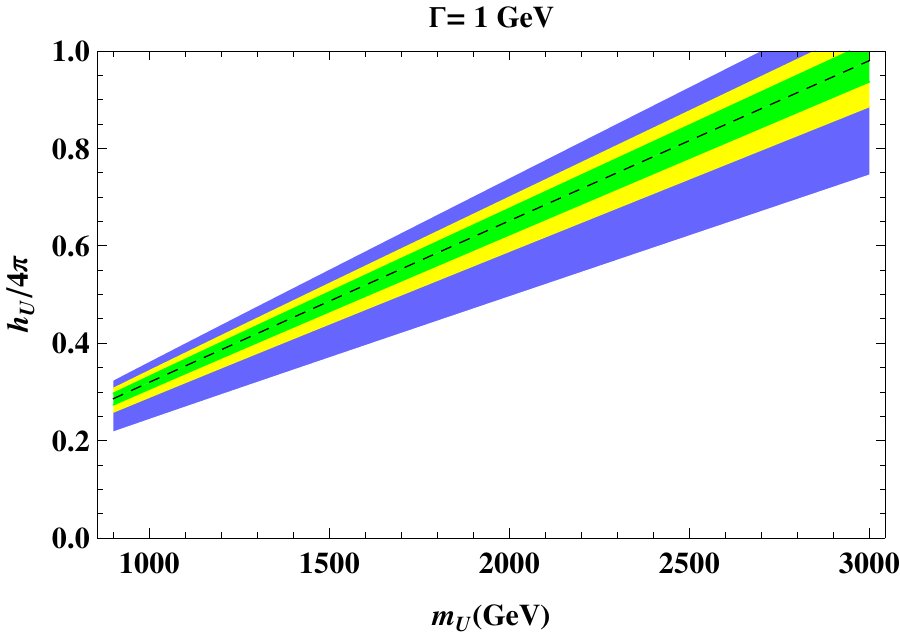}

}$\qquad$\subfloat[]{\includegraphics[scale=0.8]{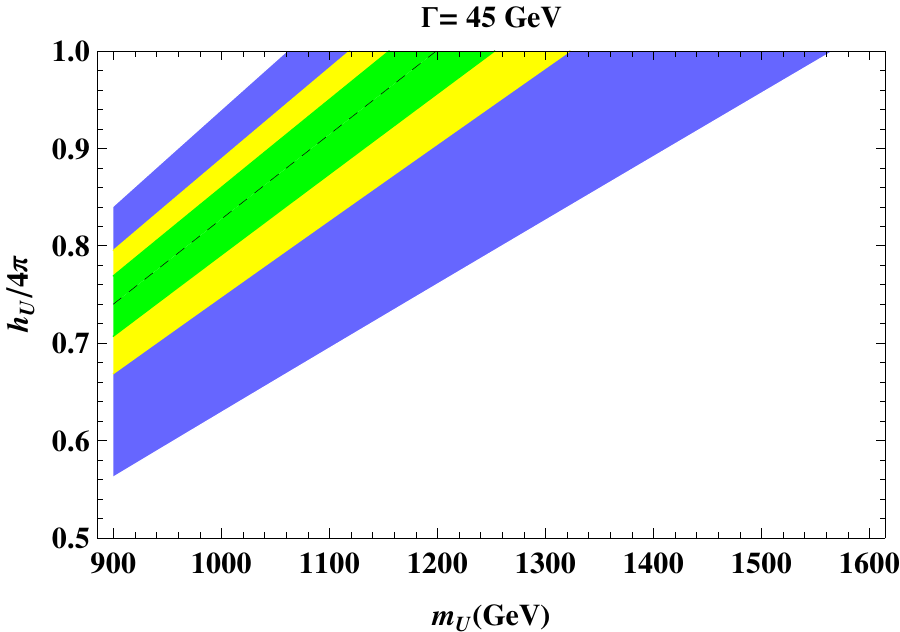}

}
\par\end{centering}

\caption{Contour plots of the production cross-section $\sigma(pp\to\xi\to\gamma\gamma)$
in femtobarns. The dashed line corresponds to the central value at
6 fb, and the shaded bands corresponds to regions at 68.3\% (green),
95.5\% (yellow) and 99.7\% (light blue) C.L. exclusion limits from
ATLAS and CMS combined data. \label{fig:Total-cross-section}}
\end{figure}

\par\end{center}

\section{Diphoton decay}

For the analysis of the diphoton decay, we take into account all the
possible decay modes of the 750 GeV candidate. Firstly, the masses
of charged Higgs bosons $H_{1,2}^{\pm}$ and $\phi_{3}^{\pm}$ are
at the TeV scale, so the decay of $\xi$ at tree level into these
charged Higgs bosons in the model is kinematically forbidden. Secondly,
when the SSB $SU(3)_{L}\rightarrow SU(2)_{L}$ takes place, $U_{L}$
does not acquire a $SU(2)_{L}$ quantum number resulting in a $SU(2)_{L}$
singlet. As a consequence of that, the decay $\xi\rightarrow WW$
is forbidden too. Thirdly, the $\xi\rightarrow ZZ$ decay is negligible
at tree-level as it is suppressed by the $Z-$mixing angle $\tan\theta_{Z}\propto{v^{2}}/{V^{2}}$.
Similarly, the $\xi\rightarrow t\overline{t}$ decay is negligible
at tree-level as the $\xi-t$ coupling is proportional to $\sin\theta_{tU}\propto v_{2}/V_{1}$.
Finally, the decay $\xi\rightarrow hh$ is strongly constrained by
ATLAS and CMS at 95\%CL \cite{ellis}. In this way, we obtain the
following total decay width for $\xi$,

\begin{equation}
\Gamma=\Gamma_{\gamma\gamma}+\Gamma_{gg}+\Gamma_{Z\gamma}+\Gamma_{ZZ}.
\end{equation}

The experimentally reported width of the resonance ranges between
0 and 100 GeV, and can be larger (`broad') or smaller (`narrow') than
the experimental resolution of about 6-10 GeV \cite{ReviewStrumia}.
The best-fit width reported by the ATLAS Collaboration is $\Gamma\sim45\,\mathrm{GeV}\sim0.06\,m_{\xi}$.
So, in view of some tension with the CMS data we use three approximations
for the decay width:
\begin{itemize}
\item A width approximation given by the experimentally reported width from
the ATLAS Collaboration $\Gamma=45$ GeV. 
\item A width approximation for a narrower resonance with $\Gamma=1$ GeV. 
\item An approximation given only by one loop contributions, $\Gamma=\Gamma_{\gamma\gamma}+\Gamma_{gg}+\Gamma_{Z\gamma}+\Gamma_{ZZ}$. 
\end{itemize}
Following \cite{Cao,Gunion} the decay rates of $\xi$ are given by

\begin{align}
\Gamma(\xi\to\gamma\gamma) & =\frac{\alpha^{2}h_{U}^{2}m_{\xi}^{3}}{512\pi^{3}m_{U}^{2}}\big|\sum_{i}N_{ci}Q_{i}^{2}F_{i}\big|^{2},\nonumber \\
\Gamma(\xi\to gg) & =\frac{\alpha_{s}^{2}h_{U}^{2}m_{\xi}^{3}}{64\pi^{3}m_{U}^{2}}\big|\sum_{i}F_{i}\big|^{2},\nonumber \\
\Gamma\left(\xi\to Z\gamma\right) & =\frac{\alpha^{2}h_{U}^{2}m_{\xi}^{3}}{64\pi^{3}m_{U}^{2}}\left(1-\dfrac{m_{Z}^{2}}{m_{\xi}^{2}}\right)^{3}\left|\frac{7\kappa^{2}c_{w}}{2s_{w}}+\frac{2}{3}\sum_{i}N_{ci}Q_{i}^{2}+\frac{\lambda^{2}}{24\pi\alpha}\right|^{2},\nonumber \\
\Gamma\left(\xi\to ZZ\right) & =\frac{\alpha^{2}h_{U}^{2}m_{\xi}^{3}}{128\pi^{3}m_{U}^{2}}\mathcal{P}\left(\dfrac{m_{Z}^{2}}{m_{\xi}^{2}}\right)\left|\frac{7\kappa^{2}c_{w}{}^{2}}{2s_{w}{}^{2}}-\frac{2}{3}\sum_{i}N_{ci}Q_{i}^{2}-\frac{\lambda^{2}}{24\pi\alpha}-\frac{\eta^{2}}{96\pi\alpha}\right|^{2}\label{eq:decays}
\end{align}

where $\mathcal{P}(x)=\sqrt{1-4x}\left(1-4x+6x^{2}\right)$ is a factor
correcting the massive final states in the decay width. Here, $h_{U}^{2}=(h_{\chi1}^{2}+h_{\psi1}^{2})/2$
, i.e, we assume the same Yukawa coupling for the three families for
simplicity and with the same mass $m_{U}$ and we have made $\kappa=\kappa^{\prime}$,
$\lambda=\lambda^{\prime}$. The functions $F_{i}$ 
\begin{equation}
F_{i}(\tau_{i})=\begin{cases}
2+3\tau_{i}+3\tau_{i}(2-\tau_{i})f(\tau_{i}) & i=1\\
-2\tau_{i}\left[1+(1-\tau_{i})f(\tau_{i})\right] & i=1/2\\
\frac{1}{2}\tau_{i}\left[1-\tau_{i}f(\tau_{i_{i}})\right] & i=0
\end{cases}
\end{equation}

\noindent are spin dependent functions for the loop factor. For $\tau_{i}>1$
the function $f(\tau_{i})$ is 
\begin{equation}
f(\tau_{i})=\left[\arcsin\left(\frac{1}{\sqrt{\tau_{i}}}\right)\right]^{2}.\label{loop-factor}
\end{equation}
with $\tau_{i}=4m_{i}^{2}/m_{\xi}^{2}$, where the masses of the particles
into the loop are $m_{i}>375$ GeV.

\begin{center}
\begin{figure}[t]
\begin{centering}
\subfloat[]{\includegraphics[scale=0.9]{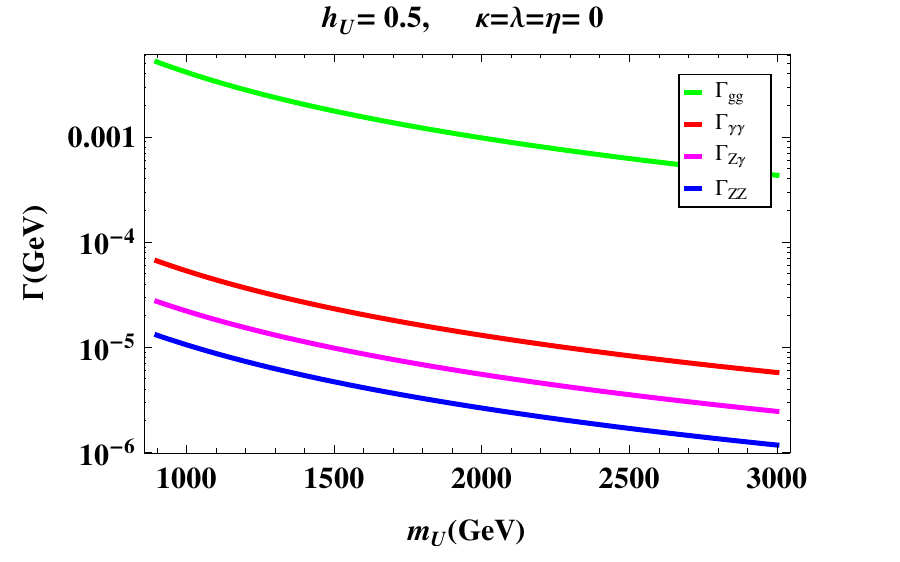}

}$\qquad$\subfloat[]{\includegraphics[scale=0.9]{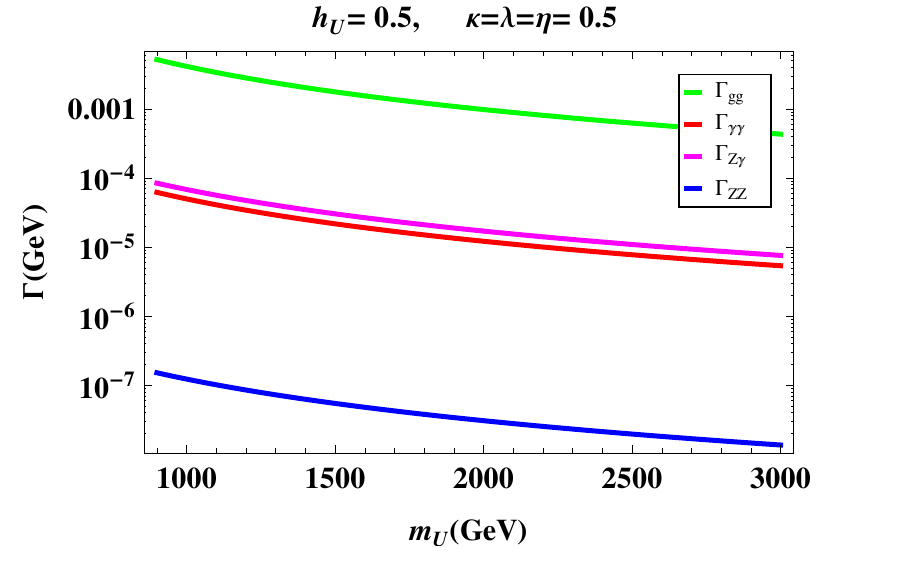}

}
\par\end{centering}

\caption{Different decay channels for the 750 GeV candidate at one loop level.
\label{fig:decays}}
\end{figure}

\par\end{center}

\subsection{Production cross section}

\noindent The total cross section $\sigma(pp\to\xi\to\gamma\gamma)$
in the narrow width approximation is given by

\begin{equation}
\sigma(pp\to\xi\to\gamma\gamma)=\frac{C_{gg}\Gamma(\xi\to gg)}{s\:m_{\xi}\Gamma}\Gamma(\xi\to\gamma\gamma).
\end{equation}
where

\begin{equation}
C_{gg}=\frac{\pi^{2}}{8}\intop_{m_{\xi/s}}^{1}\frac{dx}{x}g(x)g(m_{\xi}^{2}/sx)
\end{equation}
is the dimensionless partonic integral computed at the scale $\mu=m_{\xi}=750$
GeV and center of mass energy $\sqrt{s}=13\,\mathrm{TeV}$, obtaining
$C_{gg}=2137$ \cite{Cgg}. For the analysis we have taken the combined-rescaled
results for the cross section from CMS and ATLAS, $\sigma(pp\to\xi\to\gamma\gamma)=(2-8)\:\mathrm{fb}$
equally valid for $\sqrt{s}=8$ TeV and $C_{gg}=174$\cite{ellis}.

Fig. \ref{fig:Loop-processes-involved} shows all the possible one
loop contributions from exotic charged Higgs bosons, gauge bosons
and fermions. In the fermionic loop to $\gamma\gamma$ we take into
account the multiplicity coming from the three families, i.e., three
exotic quarks and nine exotic charged leptons. For this reason, the
contribution coming from the charged Higgs bosons is almost negligible.
We also take $m_{W_{3}^{\pm}}\sim3$ TeV according to experimental
constraints obtained by ATLAS and CMS Collaboration \cite{CMS ATLAS W' mass}.
However, for $m_{W_{3}^{\pm}}\sim3$ TeV the associated form factor
$F_{1}$ reaches its asymptotic value so the cross section dependence
on $m_{W_{3}^{\pm}}$ is suppressed. So, the production cross section
will depend only on the Yukawa coupling $h_{U}$, the mass of the
quarks $m_{U}$ and on the exotic charged lepton masses $m_{E}$,
$m_{E_{1}}$ and $m_{E_{2}}$. From the lower bound reported by the
ATLAS Collaboration searches on exotic heavy charged leptons \cite{lepton masses}
we set $m_{E}=m_{E_{1}}=m_{E_{2}}\sim600$ GeV.

\begin{center}
\begin{figure}[t]
\begin{centering}
\subfloat[]{\includegraphics[scale=0.8]{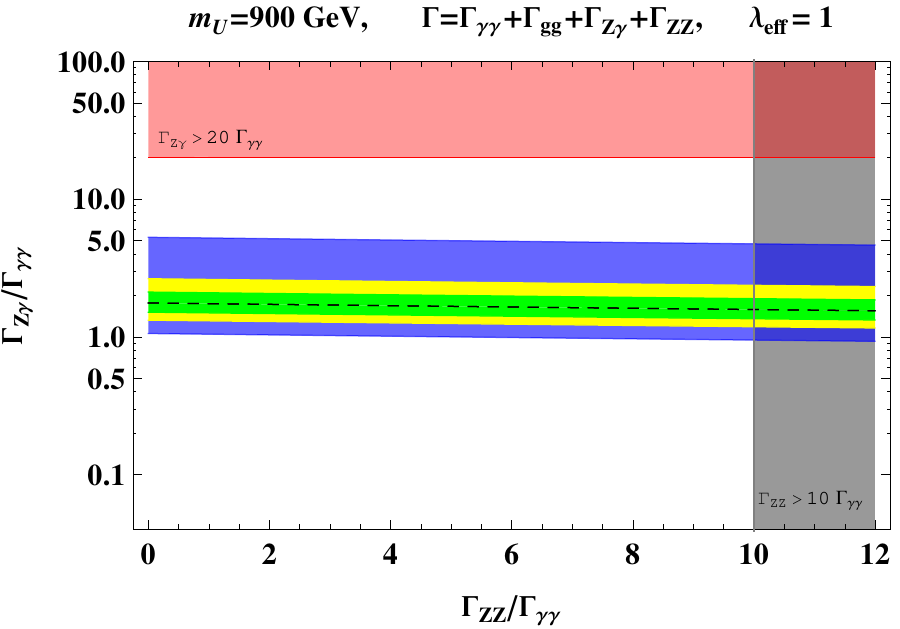}

}$\qquad$\subfloat[]{\includegraphics[scale=0.8]{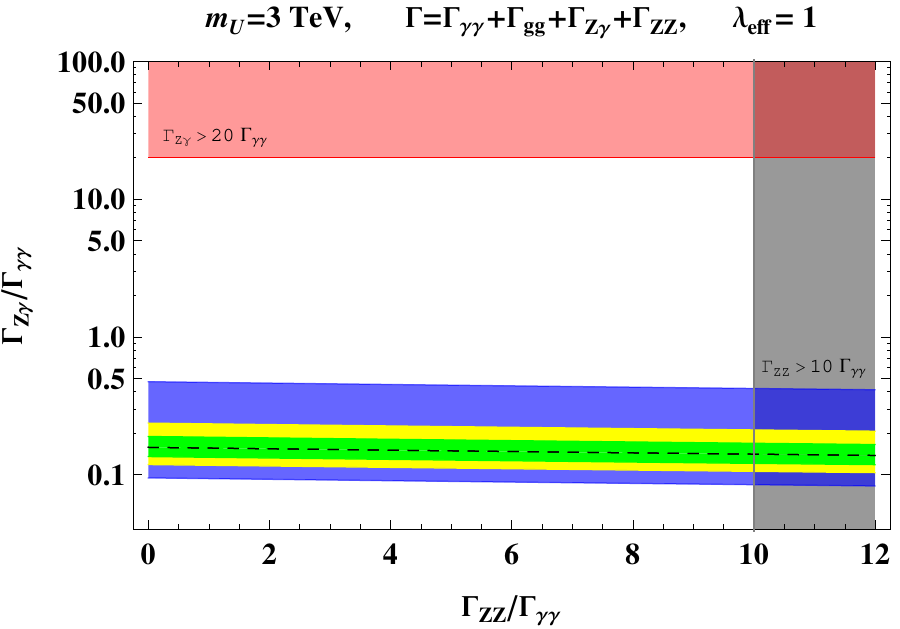}

}
\par\end{centering}

\begin{centering}
\subfloat[]{\includegraphics[scale=0.8]{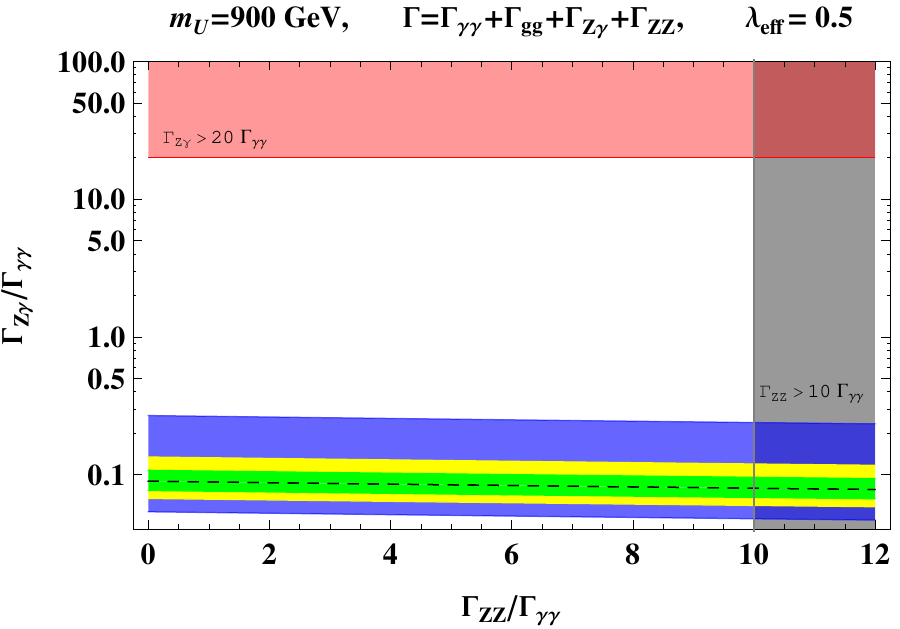}

}$\qquad$\subfloat[]{\includegraphics[scale=0.8]{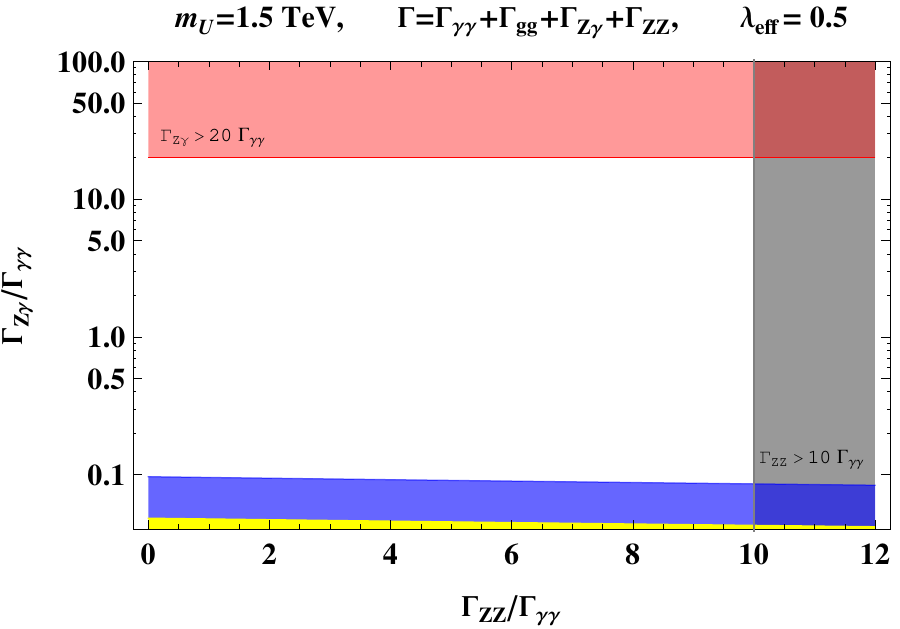}

}
\par\end{centering}

\caption{Contour plots of the production cross-section $\sigma(pp\to\xi\to\gamma\gamma)$
in femtobarns. The dashed line corresponds to the central value at
6 fb, and the shaded bands corresponds to regions at 68.3\% (green),
95.5\% (yellow) and 99.7\% (light blue) C.L. exclusion limits from
ATLAS and CMS combined data. The shaded red and gray regions are excluded.
\label{fig:exclusionplot} }
\end{figure}

\par\end{center}

Taking into account all the above conditions, we display in Fig.\ref{fig:Total-cross-section}
contour plots of the production cross-section $\sigma(pp\to\xi\to\gamma\gamma)$
as function of the up-type quark mass $m_{U}$ and the Yukawa coupling
normalized as $h_{U}/4\pi$ for $\Gamma=1$ GeV and $\Gamma=45$ GeV.
The lower bound of 900 GeV for $m_{U}$ corresponds to the reported
value in recent searches on top- and bottom-like heavy quarks from
ATLAS and CMS Collaborations \cite{quark masses} and the upper bound
of 3 TeV corresponds to the asymptotic value obtained from the fermionic
from factor $F_{1/2}$. We obtain allowed regions for both $\Gamma=1$
GeV and $\Gamma=45$ GeV widths for the scalar particle of 750 GeV
in agreement with the ATLAS and CMS Collaborations data. In Fig.\ref{fig:Total-cross-section}
(a) we obtain values for $h_{U}/4\pi$ from $0.2$ to $1.0$ and an
allowed mass region for the up-like quark from 900 GeV to 3.0 TeV
at 99.7\% CL. In Fig.\ref{fig:Total-cross-section} (b) the model
is excluded for $h_{U}/4\pi<0.6$ and $m_{U}>1.5$ TeV at 99.7\% CL.

Finally, for the case $\Gamma=\Gamma_{\gamma\gamma}+\Gamma_{gg}+\Gamma_{Z\gamma}+\Gamma_{ZZ}$,
we show in Fig. \ref{fig:decays} the different contributions in Eq.
(\ref{eq:decays}) for the decay width of $\xi$. From Fig. \ref{fig:decays}
(a), the case $\kappa=\lambda=\eta=0$ and $h_{U}=0.5$ corresponds
to pure fermionic contributions into the loops. We can see that the
contributions (ignoring the dominant $\Gamma_{gg}$) $\Gamma_{\gamma\gamma},$
$\Gamma_{Z\gamma}$, $\Gamma_{ZZ}$ have branching ratios of order
$64\%,$ $25\%$, $11\%$ respectively. On the other hand, the case
$\kappa=\lambda=\eta=0.5$ and $h_{U}=0.5$ in Fig. \ref{fig:decays}
(b), corresponds to both fermionic and bosonic contributions into
the loop with $\mathrm{BR}_{\gamma\gamma},$ $\mathrm{BR}_{Z\gamma}$,
$\mathrm{BR}_{ZZ}$ of order $43\%,$ $56\%$, $1\%$ respectively.

In this way, and taking into account current bounds on $\Gamma_{Z\gamma}/\Gamma_{\gamma\gamma}$
and $\Gamma_{ZZ}/\Gamma_{\gamma\gamma}$ \cite{Strumia}, we display
in Fig.\ref{fig:exclusionplot} contour plots of the production cross-section
$\sigma(pp\to\xi\to\gamma\gamma)$ in the $\Gamma_{Z\gamma}/\Gamma_{\gamma\gamma}$-$\Gamma_{ZZ}/\Gamma_{\gamma\gamma}$
plane. For simplicity, we have set $\lambda_{eff}\equiv\kappa=\lambda=\eta=h_{U}$
in such a way that the contour plots only depend on $m_{U}$ and $\lambda_{eff}$.
In general, for low values of $m_{U}$ the ratio $\Gamma_{Z\gamma}/\Gamma_{\gamma\gamma}$
is of order $\Gamma_{Z\gamma}/\Gamma_{\gamma\gamma}\sim1$, and for
greater values of $m_{U}$ we have $\Gamma_{Z\gamma}/\Gamma_{\gamma\gamma}<1$.
We also observe that the greater the ratio $\Gamma_{Z\gamma}/\Gamma_{\gamma\gamma}$,
the stronger the coupling $\lambda_{eff}$. However, if $\lambda_{eff}>3$
the model is completely excluded by the bound $\Gamma_{Z\gamma}<20\Gamma_{\gamma\gamma}$
for all $m_{U}$.

\section{Summary}

We have presented an anomaly-free model based on the electroweak-strong
unification group $SU(6)\otimes U(1)_{X}$, containing the $SU(3)_{C}\otimes SU(3)_{L}\otimes U(1)_{X}$
as a subgroup. We break the gauge symmetry down to $SU(3)_{C}\otimes U(1)_{Q}$
and at the same time give masses to the fermion fields in the model
in a consistent way by five Higgs fields $\Phi$, $H_{1}$, $H_{2}$
, $H_{3}$ and $H_{S}$. These Higgs fields and their VEVs set two
different mass scales: $\upsilon=256$ GeV $<<V$. From the mixing
terms between the real fields $\xi_{1}$ in $H_{1}$ and $\xi_{S}$
in $H_{S}$, we obtain two real scalar fields: our candidate for the
750 GeV signal $\xi$, and one $\xi'$ at the TeV scale. For the analysis
of the diphoton decay, we take into account all the possible decay
modes of the 750 GeV candidate considering three approximations for
the decay width: $\Gamma=1$ GeV, $\Gamma=45$ GeV and $\Gamma=\Gamma_{\gamma\gamma}+\Gamma_{gg}+\Gamma_{Z\gamma}+\Gamma_{ZZ}$.
Then, taking various simplified assumptions on the parameter space,
we show that the states $U$, $E$, $E_{1}$, $E_{2}$, $W_{3}^{\pm}$,
$H_{1,2}^{\pm}$ and $\phi_{3}^{\pm}$ into the loop can explain the
diphoton excess for each one of the width approximations according
to ATLAS and CMS bounds on all the particle masses involved and on
the decay widths $\Gamma_{Z\gamma}$ and $\Gamma_{ZZ}$.

\section*{Acknowledgment }

This work was supported by El Patrimonio Autónomo Fondo Nacional de
Financiamiento para la Ciencia, la Tecnología y la Innovación Francisco
José de Caldas programme of COLCIENCIAS in Colombia.

\end{document}